# Topological defects as relics of emergent continuous symmetry and Higgs condensation of disorder in ferroelectrics

Shi-Zeng Lin[1*], Xueyun Wang[2*], Yoshitomo Kamiya[1,3], Gia-Wei Chern[1], Fei Fan[2,4], David Fan[2,5], Brian Casas[2,6], Yue Liu[2,7], Valery Kiryukhin[2], Wojciech H. Zurek[1], Cristian D. Batista[1], and Sang-Wook Cheong[2]

[1]Theoretical Division, T-4 and CNLS, Los Alamos National Laboratory, Los Alamos, New Mexico 87545, USA

[2] Rutgers Centre for Emergent Materials and Department of Physics and Astronomy, Rutgers University, 136 Frelinghuysen Road, Piscataway, New Jersey 08854, USA

[3]iTHES Research Group and Condensed Matter Theory Laboratory, RIKEN, Wako, Saitama 351-0198, Japan

[4]Shaanxi Key Laboratory of Condensed Matter Structures and Properties, School of Science, Northwestern Polytechnical University, Xi'an 710129, China

[5]Montgomery High School, 1016 Route 601 Skillman, New Jersey 08558, USA

[6]Functional Materials Laboratory, Department of Physics, University of South Florida, Tampa 33613, USA

[7]The MOE key Laboratory of Weak-Light Nonlinear Photonics and TEDA Applied Physics School, Nankai University, Tianjin 300457, China

Corresponding Author: Sang-Wook Cheong

* These authors contribute equally

**Lars Onsager and Richard Feynman envisioned that the three-dimensional (3D) superfluid-to-normal λ transition in $^4$He occurs through the proliferation of vortices. This process should hold for every phase transition in the same universality class. The role of topological defects in symmetry-breaking phase transitions has become a**

prime topic in cosmology and high-temperature superconductivity, even though direct imaging of these defects is challenging. Here we show that the U(1) continuous symmetry that emerges at the ferroelectric critical point of multiferroic hexagonal manganites leads to a similar proliferation of vortices. Moreover, the disorder field (vortices) is coupled to an emergent U(1) gauge field, which becomes massive by means of the Higgs mechanism when vortices condense (span the whole system) upon heating above the ferroelectric transition temperature. Direct imaging of the vortex network in hexagonal manganites offers unique experimental access to this dual description of the ferroelectric transition, while enabling tests of the Kibble-Zurek mechanism.

Phase transitions are among the most fascinating phenomena of nature. Understanding their mechanisms is one of the foremost challenges of modern physics. In a nutshell, the goal is to understand how the "generalized rigidity" of the ordered phase emerges through a spontaneous symmetry breaking upon cooling across the critical temperature $T_c$[1]. An attractive aspect of these transitions is their universal behavior that makes them independent of microscopic details. The universality class is determined by a few fundamental properties including symmetry, range of interactions, dimensionality and number of components of the order parameter. In addition, the Wilson-Fisher[2] paradigm gave birth to another deep concept known as *emergent symmetry*. The basic idea is that the effective action that describes the long wavelength fluctuations near the critical point may have more symmetries than the original microscopic action. This notion implies that systems such as magnets or ferroelectrics, which only possess *discrete* symmetries at a microscopic level, may have an emergent *continuous* symmetry at a critical point, as was demonstrated by Jose, Kadanoff, Kirkpatrick and Nelson in their seminal work[3].

As envisioned by Onsager and Feynman[4,5], the restoration of a continuous U(1) symmetry, like the superfluid to normal transition of $^4$He, can occur via proliferation of vortices. The role of these topological defects in symmetry-breaking phase transitions is now a prime topic in different areas of physics, such as cosmology[6,7] and high-$T_c$ superconductivity[8-12], even though they are difficult to observe[13]. The potential of having emergent continuous symmetries in magnets or ferroelectrics with discrete microscopic symmetries opens the possibility of observing a similar proliferation of vortices in insulating materials. While the emergence of continuous symmetries from discrete variables is theoretically established, the same is not true at the experimental level. Among other reasons, it is always challenging to measure critical exponents with the required resolution to distinguish between discrete and continuous symmetry breaking. Here we demonstrate the emergence of a continuous U(1) symmetry at the ferroelectric transition of the hexagonal manganites RMnO$_3$ (R=Y, Ho, ... Lu, Sc) by directly measuring the vortices or *disorder field*, instead of addressing the order parameter field. An emergent U(1) symmetry implies that the critical point belongs to the XY universality class, which is the class of the superfluid transition of a neutral system such as $^4$He. Therefore, analogous with the case of superfluid $^4$He, the transition must be driven by a proliferation of vortices spanning the whole system above $T_c$. In this dual description, based on the disorder field of topological vortices instead of the order parameter field[14-16] (see Box 1), the phase transition is described as a condensation of the disorder field which is coupled to a gauge field[10,14]. Upon heating across $T_c$, the vortex condensation makes the gauge field massive via the Higgs mechanism. Consequently, the vortex-vortex interaction becomes screened above $T_c$, instead of the Biot-Savart interaction that characterizes the Coulomb phase below $T_c$. Fig. 1 (a,b) shows the same phase diagram from two different viewpoints with the order and disorder fields. We will see below that the possibility of freezing vortices in the hexagonal RMnO$_3$ provides a unique opportunity to experimentally access the dual theory and the Higgs condensation of disorder in insulating materials.

The dynamics of symmetry breaking in phase transitions is another fascinating phenomenon that can be tested in RMnO$_3$. Its salient features are captured by the Kibble-Zurek mechanism (KZM), which combines cosmological motivations with information about the near-critical behavior. Symmetry breaking is thought to be responsible for emergence of the familiar fundamental interactions from the unified field theory at GUT temperatures of ~$10^{15}$ GeV in the Universe cooling after Big Bang. As Kibble[17] noted, relativistic causality limits the size of domains that can coordinate the choice of broken symmetry in the nascent Universe. This results in a random selection of local broken symmetry, and can lead to creation of topological defects (e.g., monopoles or cosmic strings) that influence evolution of the Universe.

Because phase transitions are ubiquitous, their dynamics can be investigated experimentally. Although relativistic causality is no longer a useful constraint in the laboratory, cosmological motivations can be combined with the scaling relations in the near-critical regime of second order phase transitions to estimate the density of topological defects as a function of the quench rate[18]. This combination defines the Kibble-Zurek mechanism[7,19].

Topological defects are not only clearly visible in RMnO$_3$, but (in contrast to e.g. superfluids) they are immobilized by the structure of the material that solidifies below the critical point. Consequently, they can be seen and counted at leisure, long after the transition. Moreover, in contrast to other systems used as a test bed for KZM[20-27], the quench timescale $\tau_Q$ can be varied over orders of magnitude. This is important, as densities of defects predicted by KZM often scale as $\tau_Q$ to small fractional power.

**Model**

The onset of ferroelectricity in RMnO$_3$ ($T_c$'s of YMnO$_3$, ErMnO$_3$, TmMnO$_3$, LuMnO$_3$ ≈ 1250, 1403, 1523, 1672 K, respectively) is triggered by a structural instability called trimerization and a subsequent ionic displacement with a net electric dipole moment. While trimerization breaks the Z$_3$ symmetry of the hexagonal lattice, the subsequent distortion breaks an additional Z$_2$ symmetry (sign of the electric polarization along the *c* axis, perpendicular to the hexagonal plane)[28-33]. The transition then breaks a Z$_3$ ×Z$_2$ symmetry group and it is described by an effective *p* = 6 clock model,

$$\mathcal{H} = J \sum_{<j,l>} \cos(\varphi_j - \varphi_l) + J' \sum_{<j,l>} \cos(\varphi_j - \varphi_l), \qquad (1)$$

known to exhibit an emergent U(1) symmetry at the critical point[3]. The phase $\varphi_j = n\pi/3$ takes six possible values when the integer *n* runs between 0 and 5. The six minima of this Hamiltonian can be labeled by the even hours on the clock face, which justifies the name of the model. The electric polarization P$_j \propto \cos(3\varphi_j)$ is perpendicular to the triangular layers. The trimerization parameter is described by a 2D vector **T**$_j \propto (\cos(2\varphi_j), \sin(2\varphi_j))$ with three possible orientations will be denoted by (α, β, γ). *J* and *J'* are effective coupling constants between nearest-neighbor variables *j,l* on the same triangular layer and on adjacent layers, respectively. Because the six-fold anisotropy becomes relevant away from the critical point, the usual U(1) vortices of the XY model are replaced by the Z$_6$ vortices (Note that Z$_6$ does not refer to the homotopy group in this context, but to the fact that these vortices are made of six domains.)that are observed in the experiments[28-33]. The formation of these 6-state vortices, shown in Fig. 1(c), originates from the cyclic arrangement of 6 interlocked structural antiphase ($\alpha$, $\beta$, $\gamma$) and ferroelectric (+/−) ground states. According to Eq. (1), there are only two low-energy cyclic arrangements, which correspond to the vortex ($\alpha^+$, $\beta^-$, $\gamma^+$, $\alpha^-$, $\beta^+$, $\gamma^-$) and the anti-vortex ($\alpha^+$, $\gamma^-$, $\beta^+$, $\alpha^-$, $\gamma^+$, $\beta^-$) in agreement with the experimental observations[28-33], see Fig. 1(c). Any other cyclic arrangement contains domain walls with higher energy ($|\Delta\varphi| > \pi/3$).

As we explained above, the six-fold anisotropy of the clock model is dangerously irrelevant[34,35] i.e., the critical point belongs to the XY universality class but the discreteness becomes relevant away from $T_c$. Because the critical region around $T_c$ can be described by the same Ginzburg-Landau $\phi^4$ theory that describes the transition of a neutral *superfluid* (see Box 1), the ferroelectric order parameter is also destroyed by proliferation of ($Z_6$) vortices. According to the dual theory (see Fig. 1 and Box 1), a finite fraction of vortex lines span the whole system above $T_c$, i.e., the corresponding vortex cores connect opposite surfaces of the sample. Therefore, it is natural to conjecture that the 6-state vortex domain structures observed on the surface of $RMnO_3$ correspond to transverse cuts of these vortex cores. This conjecture is substantiated by the Piezo-response Force Microscope (PFM) 3D images of vortex cores in the hexagonal-$LuMnO_3$ crystal (See Supplementary Information section 1 for the method). These images are obtained after a sequence of three steps: a) heating a hexagonal $LuMnO_3$ single crystal up to $T = 1723$K ($> T_c \approx 1672$ K), b) keeping the temperature constant for 30 minutes, and c) cooling the specimen down to room temperature. The emergence of a vortex domain pattern on the *a-b* surface of the crystal is revealed by chemical etching. To obtain a 3D picture of vortex cores (i.e. a depth profiling of vortex domain patterns), the sample is polished along the *c* axis and PFM images of the same region are taken for different depths (see Fig. 2(a)). Enlarged areas exhibiting the depth evolution of vortex cores in red and green boxed regions are shown in Fig. 2 (b) and (c), respectively. The large spatial extension of vortex cores is evident in the figure. The schematics of the vortex networks shown in Figs. 2(d) and (e) are obtained by depth profiling of the 2D vortex cores.

To simulate this phenomenon, we ran Monte Carlo (MC) simulations of Eq. (1) based on a *local* update Metropolis algorithm because the microscopic dynamics of $RMnO_3$ is expected to be local (See Supplementary Information section 2 for the simulation details). The lattice size is $L^2 \times L_z$ with $L = 192\,a$ and $L_z = 96\,a$ ($a$ is the lattice parameter). We used periodic boundary conditions in the xy plane and open boundary condition along the z direction. The critical temperature obtained for $J'=J$ is $T_c \approx 3.03\,J$. The vortices shown in Fig. 2(f) were obtained after annealing from an initial temperature $T_i = 6.0\,J$ down to $T_f = 0$ with a rate $\Delta T=0.005J$ per MC sweep (MCS). To verify that they reproduce the experimental observation (Fig. 2a), we compared the distributions of point-like vortices and anti-vortices on a given layer (see inset of Fig. 3a). We define the vortex-antivortex pair correlation function as $G(r) = <n_s(0)n_s(r)>$, where $n_s(r) = \sum_\alpha q_\alpha \delta(r - r_\alpha)$ is signed defect density, $q_\alpha=1$ (-1) for vortices (antivortices), and $r_\alpha$ is the defect position. The distances are scaled by the average defect separation $r_{av}$, making G(r) dimensionless. Both the overall domain patterns and the G(r) obtained from the MC simulations (Fig. 3a) reproduce well the results obtained from applying the same analysis to the 2D experimental images (Fig. 3b) of three different hexagonal manganites with different vortex densities: ErMnO$_3$ (0.0046 μm$^{-2}$), YbMnO$_3$ (0.16 μm$^{-2}$), and YMnO$_3$ (0.19 μm$^{-2}$) (See Supplementary Information section 3 for details).

In the dual description, the disorder field (i.e. vortices and antivortices) condenses upon heating across $T_c$, and this disorder condensation makes the gauge field massive. This Higgs transition of the vortex field has direct consequences on the nonequilibrium symmetry breaking process that takes place when the temperature is lowered at a finite rate from an initial temperature $T_i$ close to $T_c$. Vortices that span the whole system disappear at a much slower rate. Consequently, the final state must be very different depending on whether the initial temperature $T_i$ is lower or higher than $T_c$. If $T_i < T_c$, all the vortices form as loops of a finite size (smaller than the system size, with the size distribution set by the Boltzmann factor) implying that they should shrink and disappear upon cooling. On the contrary, if $T_i > T_c$, a significant (~70%) fraction of the vortex network comes as an "infinite string" that spans the whole system[36] and can be expected to survive upon cooling. Our experiments and MC simulations confirm this analysis. Fig. 4(a) and (b), showing Atomic Force Microscope (AFM) images on LuMnO$_3$ with $T_c$=1672 K annealed from 1698 K and 1673 K, demonstrate the presence of vortices/antivortices, indicating a finite vortex density for $T_i > T_c$. When LuMnO$_3$ is annealed from temperatures lower than $T_c$, annular domain patterns, high-density wavy stripes, and low-density straight stripes are found with annealing $T$'s at 1671 K, 1633K and 1593 K, respectively. The optical microscope images are shown in Fig. 4(c)-(e). The trend is evident: many vortices remain in the final state when $T_i > T_c$. In contrast, the final state for $T_i < T_c$ consists of annular patterns or straight stripes, and contains no vortices or antivortices (Stripes are expected to form due to the long-range dipolar interactions, which are present in the real system, but are not included in our model). Indeed, our experimental results and numerical simulations indicate that this discontinuous change in the dynamics of the vortex field can be used to determine $T_c$ in a very precise way (±1 K/1672 K=±0.06%).

## Kibble-Zurek Mechanism

In the previous section we have seen that the equilibrium properties of RMnO$_3$ ferroelectrics are well described by a Z$_6$ clock model and that the relevant degrees of freedom freeze below a certain temperature $T<T_c$. This combination provides an ideal setting for testing the KZM. The main difference between the cosmological and laboratory settings is that now the relaxation time and coherence length (and speed of the relevant sound rather than the speed of light) determine the sonic horizon -- the linear size $\hat{\xi}$ of regions that can break symmetry in step. The basic idea[18] is to compare the relaxation time $\tau$ with the timescale of change of the key parameter [here, relative temperature $\varepsilon = (T - T_c)/T_c$]. We assume $\varepsilon = t/\tau_Q$, where $\tau_Q$ is the quench time. The relaxation time $\tau(\varepsilon) = \tau_0/|\varepsilon|^{\nu z}$ (where $\nu$ and $z$ are spatial and dynamical critical exponents, and $\tau_0$ is a timescale set by microphysics) determines the reaction time of the order parameter. Relaxation characterized by $\tau(\varepsilon)$ is faster than $|\varepsilon/\dot{\varepsilon}| = |t|$ outside interval $\hat{t} = \left(\tau_0 \tau_Q^{z\nu}\right)^{1/(1+\nu z)}$ around the transition, so the system can quasi-adiabatically follow the change imposed by the quench. This instant is determined by the equation[18]:

$$\tau(\varepsilon(\hat{t})) = |\varepsilon/\dot{\varepsilon}| = \hat{t}$$

The system will cease to keep up with the imposed change at time $\hat{t}$ before reaching the critical point, while its reflexes are recovered at time $\hat{t}$ (i.e., when $\hat{\varepsilon} = (\tau_0/\tau_Q)^{1/(1+\nu z)}$) after the transition. Thus, broken symmetry is chosen by fluctuations when their coherence length is[18]:

$$\hat{\xi} = \frac{\xi_0}{|\hat{\varepsilon}|^\nu} = \xi_0 (\tau_Q/\tau_0)^{\nu/(1+\nu z)}$$

The choice of broken symmetry is random within fluctuating domains of this size. Topological defects are then expected to form with the density of one defect fragment per domain. Thus, the scaling of $\hat{\xi}$ with quench rate set by the universality class of the transition translates into the scaling of the defect density. This prediction has been verified for the 3D XY model[37]. Indeed, even the KZM predictions of the actual density (and not just its scaling) are close to these observed[38]. Our experimental results (see Supplementary Information section 4) for RMnO$_3$, as well as our simulations (See Supplementary Information section 2 for the simulation details) of $\mathcal{H}$ of Eq. (1), confirm this prediction and corroborate KZM (see Fig. 5). The obtained exponent of ~0.59 is very close to the value $2\nu/(1+\nu z) \cong 0.57$ that is expected for a 3D XY fixed point: $\nu=0.67155(27)$[39] and $z \cong 2$.[40] We emphasize that the 3D XY fixed point is a consequence of the $Z_6$ symmetry of RMnO$_3$ compounds (the $Z_6$ anisotropy is dangerously irrelevant at $T=T_c$[34,35]). Moreover, we verify the KZM for rapid quenches (where Ref.[38] observed an unexpected decrease of defect density they termed "anti-KZM").

By using the above estimate of one $\hat{\xi}$ defect fragment per volume of the domain of that linear size, one can estimate the defect density as a function of quench rate and of $\tau_0$ and $\xi_0$ – two dimensionful constants that characterize the system – and its universality class given by the spatial and dynamical critical exponents $\nu$ and $z$.

The essence of Kibble-Zurek mechanism (KZM) is the randomness of the choices of broken symmetry in domains of size $\hat{\xi}$. This randomness – in addition to defect density – predicts[18] scaling of the winding number $W$ subtended by a contour $C$. The winding number is the net topological charge: $W = n_+ - n_-$, the difference of the numbers $n_+$ and $n_-$ of vortices and antivortices inside $C$. If these charges were assigned at random, typical net charge would be proportional to the square root of their total number, $n = n_+ + n_-$, inside $C$, so it would scale as a square root of the area $A$ inside $C$. Therefore, for contours of a fixed shape, it would scale as the length of the contour, $W \propto \sqrt{A} \propto C$.

According to the KZM, $W$ is set by the winding of the phase along $C$. In our clock model, broken symmetry phases correspond to even hours on the clock face. $W$ is then the "number of days" elapsed along the contour $C$. As choices of even hours (phases) are random in $\hat{\xi}$-sized domains, the typical net winding number $W$ scales as $\sqrt{C/\hat{\xi}}$ – it is proportional to the square root of the number of steps.

This scaling with the *square root* of the circumference can be tested by finding the net charge of vortices inside $C$. The results are shown in Fig 6. The typical winding number (characterized either by the average absolute value $\langle |W| \rangle$ or the dispersion $\sqrt{\langle W^2 \rangle}$) indeed scales like $\sqrt{C/\hat{\xi}}$, as long as $C > \hat{\xi}$.

This scaling dependence changes when the magnitude of $W$ falls below 1. Moreover, the scalings of $\langle |W| \rangle$ and of the dispersion $\sqrt{\langle W^2 \rangle}$ diverge in this regime. This may seem surprising, but it is actually predicted by the KZM[41]: $|W|<1$ occurs when $C < \hat{\xi}$, i.e., $C$ normally contains a single defect or none. In this case $\langle |W| \rangle \approx p_+ + p_- = p_{DEFECT}$, while $\sqrt{\langle W^2 \rangle} = \sqrt{p_{DEFECT}}$ in terms of probabilities. Moreover, the probability $p_{DEFECT}$ of finding a defect inside $C$ is proportional to area $A$ subtended by $C$, accounting for both the change and divergence of the scalings of $\langle |W| \rangle$ and $\sqrt{\langle W^2 \rangle}$ seen in Fig. 6a.

Further evidence of the KZM is found in the scaling of $\langle |W| \rangle$ and $\sqrt{\langle W^2 \rangle}$ with the deformation of the shape of the contour (and the consequent changes of the area $A$ inside). Fig. 6 b,c shows that, as long as the size of the contour is large compared to $\hat{\xi}$, the winding number depends only on its length, and not on the area enclosed by the loop. However, as expected, the area becomes important when the number of defects falls below 1 and the scalings of $\langle |W| \rangle$ and $\sqrt{\langle W^2 \rangle}$ steepen and diverge.

This prediction can be motivated[41] by using a simple model where defects of opposite charge appear in pairs and both their size and typical separations are given by $\hat{\xi}$. This pairing of vortices and antivortices is an *ad hoc* model, although it can be motivated by considering the three – dimensional geometry of vortex lines and their relation to the vortices that appear as a 2D plane intercepts a 3D sample (see e.g. Fig. 2). Moreover, pairing of the oppositely charged defects is obviously consistent[13,42] with the correlation functions seen in Fig. 3.

## Conclusions

The observation of the Kibble-Zurek mechanism in a non-equilibrium phase transition with the scaling of defects density determined by the exponents of the 3D XY model provides further confirmation of the emergent U(1) at the critical point of the structural transition. It is then natural to expect a proliferation of vortices right above the transition. However, because the symmetry remains discrete at temperatures away from the critical point, the continuous U(1) vortices that appear in superfluid systems are replaced by the discrete $Z_6$ vortices observed in hexagonal $RMnO_3$ at room temperature. This novel realization of a 3D XY transition has a unique advantage relative to superfluids: the possibility of freezing vortices in hexagonal RMnO3 provides a unique opportunity to study the disorder field (dual theory) and measure the experimental consequences of the Higgs mechanism that arises from its condensation. In particular, geometric properties of the vortex field of $RMnO_3$ can be related to the critical exponents of the dual abelian gauge theory that describes the disorder field (see Box 1). For instance, the critical exponent that controls how the effective line tension of vortices vanishes when $T_c$ is approached from below is $\gamma = \nu_\psi(2 - \eta_\psi)$, where $\nu_\psi$ and $\eta_\psi$ are the critical exponents of the dual or *disorder matter field*[43]. Because the dual field is described by the same abelian gauge theory that describes the *order parameter field* of a charged superfluid, measuring the critical behavior of vortices in $RMnO_3$ would provide information about the critical exponents of a charged superfluid, such as a strongly type II superconductor.

The implications of our discussion go far beyond multiferroics. For example, experiments[44,45] with rapid cooling of superconducting loops reported that the frequency of trapping a single flux quantum inside scaled with a power which was *four times* that predicted for large loops. We have seen such steepening in $\langle |W| \rangle \approx p_+ + p_- = p_{DEFECT}$ that is indeed, in the $p_{DEFECT} < 1$ regime, set by $p_{DEFECT}$, and thus proportional to $A/\hat{\xi}^2$. Thus, an explanation (based on the fabrication problems) of the apparent divergence between what was thought to be KZM predictions and experiment put forward before turns out to be unnecessary. Instead of the doubling, (presumably based on the expected behavior[46] of $\sqrt{\langle W^2 \rangle}$), the scaling of the frequency of trapping a flux quantum with quench rate quadruples (see Fig. 6). Such quadrupling of $p_{DEFECT}$ and $\langle |W| \rangle$, predicted by the KZM, (and seen before[44,45]) is hereby confirmed experimentally (Fig. 6).

Last but not least, the confirmation of the KZM in a 3D XY critical point (the same universality class as the λ transition in $^4$He) suggests that the failure to detect[47] KZM vortices in $^4$He quench experiments may be due to their rapid annihilation combined with the inability to measure their density right after the quench. Ferroelectrics bypass this problem by immobilizing vortices in the matrix of the material soon after the quench, preventing their annihilation.

**Supplementary Information** is linked to the online version of the paper at www.nature.com/

**Additional information**

The authors declare no competing financial interests. Correspondence and requests for materials should be addressed to S.-W. C. (sangc@physics.rutgers.edu).

**Acknowledgements**

We thank A. del Campo and Vivien Zapf for stimulating discussions. This project was in part supported by DOE under the LDRD program at the Los Alamos National Laboratory, and the work at Rutgers University was supported by the DOE under Grant No. DE-FG02-07ER46382. Y.K. acknowledges the financial support by the RIKEN iTHES Project. The work was also supported by China Scholarship Council.

**Author contributions**

S.-W. C. designed and supervised the experimental part. W. H. Z and C. D. B. discussed the simulations and experiments, and wrote the Kibble-Zurek Mechanism part. X. W. carried out annealing experiments, AFM and PFM work. F. F. performed PFM work. D. F., B. C. and Y. L. analyzed vortex-antivortex optical images, and V. K. calculated experimental correlation functions. S.-Z. L., Y. K. and G.-W. C. simulate the theoretical results, and S.-Z. L., X. W., S.-W. C., W. H. Z., C. D. B. and V. K. co-wrote the paper, and all authors discussed the results.

**FIG. 1 | Dual description of a phase transition with $Z_2 \times Z_3$ symmetry.** The phase transition can be described in terms of (a) order field $\Phi$ or (b) disorder field $\Psi$. The local order parameter $\Phi$ takes six values, represented by the even hours in the clock dials in (a). They correspond to the six multiferroic states or domains $\alpha^+$ through $\gamma^-$ distinguished by the polarization direction (+ or -), and the trimerization phase ($\alpha$, $\beta$, $\gamma$), as described in the text. The multiferroic $Z_2 \times Z_3$ vortices are line defects where the six domains meet with each other, as shown in (c). Continuous U(1) symmetry emerges from $Z_2 \times Z_3$ order parameter at the critical temperature. The disordered phase above $T_c$ can be described as a condensation of the disorder field $\Psi$ signaled by the proliferation of vortex lines spanning the whole system (yellow lines). Only quickly fluctuating closed vortex loops (red lines) are present for $T<T_c$. The Higgs and Coulomb phases of the disorder field are described in Box 1.

**FIG. 2 | 3D picture of vortex cores: a depth profiling of vortex domain patterns.** (a) Evolution of the $Z_2 \times Z_3$ ferroelectric domains on the polished surface of a hexagonal-LuMnO$_3$ crystal, which was consecutively thinned down. The depth of each polished surface from the original surface is shown in micro-meters. Layer 1 is an atomic force microscope image on the unpolished, but chemically etched surface. Layers 2-6 are piezo-response force microscope images on the same area with different depths. The dark and light regions correspond to the domains with the opposite electric polarizations. (b) and (c) are enlarged areas exhibiting the evolution of the vortex and antivortex cores (labeled as blue and red solid dots, respectively) in red-boxed regions and green boxed regions. Vortex-antivortex pairs can be observed. The structural phases forming such pairs are identified in Fig. 3(b). Panels (d) and (e) depict the obtained depth profiles of the vortex cores for the regions shown in (b) and (c), respectively. (f) Vortex loops and lines spanning the whole system obtained from our Monte Carlo simulations of the 3D clock model of Eq. (1) at $T > T_c$.

**FIG. 3 |Vortex-antivortex pair correlation function.** (a) Theoretical vortex-antivortex pair correlation function G(r) for the configuration obtained on the surface of the system after annealing from $T_i = 6.0\ J$ to $T_f = 0.0\ J$ with a rate $\Delta T = 0.001\ J$ per MC sweep. (b) Measured G(r) for three different samples with different vortex densities: ErMnO$_3$ (green),

YMnO$_3$ (black), and YbMnO$_3$ (red). The distances are scaled by the average defect separation r$_{av}$, making the shown quantities dimensionless. Error bars are from counting statistics, and represent one standard deviation. The insets show parts of the domain patterns obtained in the calculations (a), and experimentally measured by AFM in YbMnO$_3$ (b). The blowups show vortex-antivortex pairs in small regions of these patterns, with the structural phases labeled.

**FIG. 4 | Domain patterns for different initial annealing temperatures $T_i$ (above, close, and below $T_c$).** Atomic Force Microscope (AFM) images (a,b), and optical images (c-e) of LuMnO$_3$ ($T_c$=1672 K), with $T_i$ indicated in each panel. Adjacent areas of different colors correspond to the domains with the opposite polarizations. Vortices are found only for $T_i>T_c$, while stripe and annular domain patterns are observed for $T_i<T_c$. This is illustrated by schematic blowups in panel (f), showing vortices for $T_i$=1673 (one degree above $T_c$), and annular patterns for $T_i$=1671 K (one degree below $T_c$). The plots in (f) correspond to the data in the green and red boxes in panels (b) and (c).

**FIG. 5 | Dependence of vortex density n$_v$ on cooling rate.** (a) Experimental vortex density in the final state as a function of the cooling rate. TmMnO$_3$ ($T_c \approx$1523 K) samples with the cooling rates ranging from 2 K/h to 12000 K/h, and ErMnO$_3$ ($T_c \approx$1403 K) with the cooling rates from 0.5 K/h to 300 K/h[37] were measured. The vortex density as a function of cooling rate is consistent with a power law dependence with exponent 0.59 (full lines), that is obtained from our MC simulations shown in (b) for a final temperature of 0.92 $T_c$ (black dots), as well as with the prediction of ~0.57 that follows from the Kibble-Zurek mechanism. In (b), the cooling speed is given in inverse MC sweep (arbitrary units), n$_{eq}$ is the density of the thermally excited vortices subtracted to reveal the KZM scaling (see Supplementary Information, Section 2), and *a* is the lattice parameter.

**Fig. 6 | Winding numbers for KZM defects.** Absolute winding numbers $\langle|W|\rangle$ and their dispersions $\sqrt{\langle W^2 \rangle}$ obtained from the coordinates of ~4100 defects in YMnO$_3$ crystal (see Supplementary Information Section 3), as functions of the average number of defects <n> inside the contour $C$. Typical winding numbers $W = n_+ - n_-$ inside $C$ are predicted by KZM[18,41]. Their scaling follows from the key idea that local random choices of broken

symmetry in domains of size $\sim \hat{\xi}$ determine defect locations. For contours with circumference $C > \hat{\xi}$ that contain many defects (large average $n = n_+ + n_-$) KZM predicts that $\langle |W| \rangle$ and $\sqrt{\langle W^2 \rangle}$ depend only on $C$, and vary as $\sqrt{C/\hat{\xi}}$, independently of its shape, area $A$, or the average $<n> \sim A$ inside $C$. This may seem surprising, for if defect charges were random, one would expect winding number (the mismatch between vortices and antivortices inside $C$) to vary as $<n>^{1/2} \sim A^{1/2} \sim C$. KZM prediction is confirmed in (a) for randomly placed contours of a fixed shape (here squares, like the green one in Fig. 4b). For a fixed shape $C \sim <n>^{1/2}$ and large $<n>$ typical winding number scaling $|W| \sim \sqrt{C/\hat{\xi}}$ result in $\langle |W| \rangle$ and $\sqrt{\langle W^2 \rangle}$ proportional to $<n>^{1/4}$. By contrast, when $<n> < 1$, there is usually at most one defect inside $C$, so $W$ can be only 0, +1, or -1, so $\langle |W| \rangle$ is proportional to probability $p$ of finding a defect. Moreover, $p \sim A$, so now $\langle |W| \rangle \sim A \sim <n>$ -- typical winding numbers depend on the area $A$ inside $C$. However, dispersion is proportional to $p^{1/2}$, so $\sqrt{\langle W^2 \rangle} \sim C \sim <n>^{1/2}$. Thus, when $<n> < 1$ scaling of average $\langle |W| \rangle$ and $\sqrt{\langle W^2 \rangle}$ differ[41]. This is also seen in panel (a). Panel (b) shows $\langle |W| \rangle$ for contours of the same circumference, but with different shapes and, hence, areas that differ by a factor of ~3. As expected, $\langle |W| \rangle$ depends on $C^{1/2}$ for large $<n>$, but on $A \sim <n> \sim C^2$ for fractional $<n>$. Panel (c) shows the same data as (b) redrawn as a function of $<n>$. In all panels, solid lines show power laws predicted for $\langle |W| \rangle$ and $\sqrt{\langle W^2 \rangle}$ by KZM[41]. $C$ is normalized so that a square of circumference $C = 4$ ($A = 1$) contains one defect on average. KZM prediction[18,41] for $W$, based on randomness of broken symmetry choice in domains of size $\sim \hat{\xi}$, is thus verified in all cases.

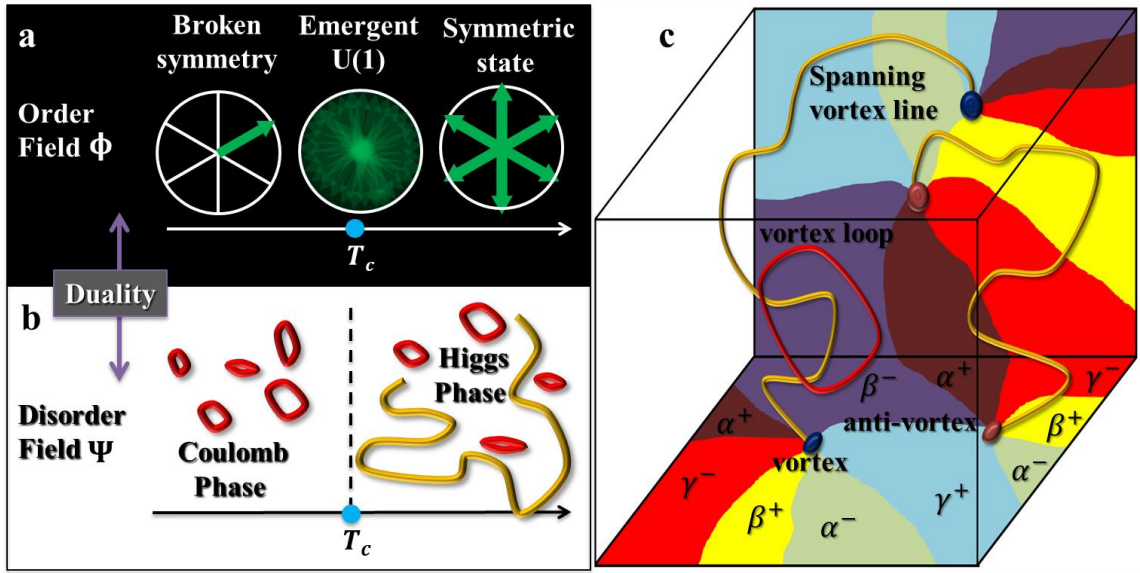

Fig. 1
20

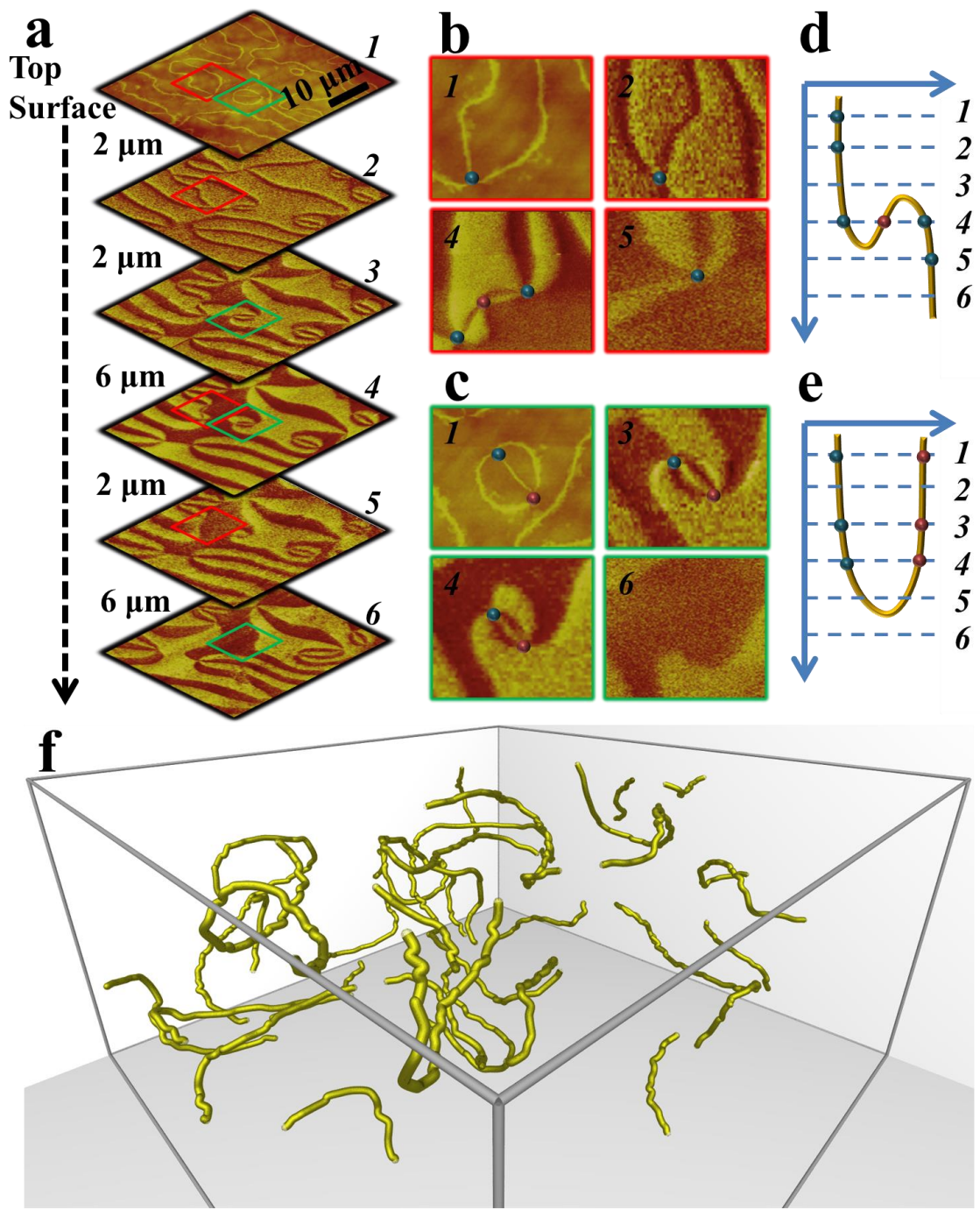

**Fig. 2**

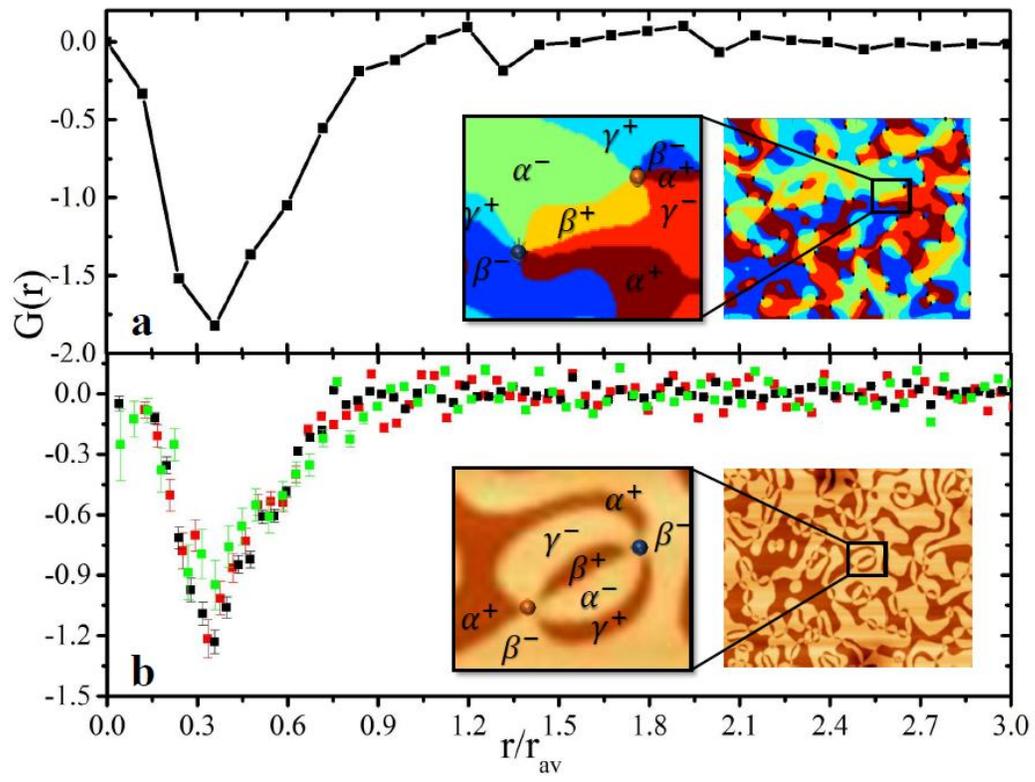

Fig. 3

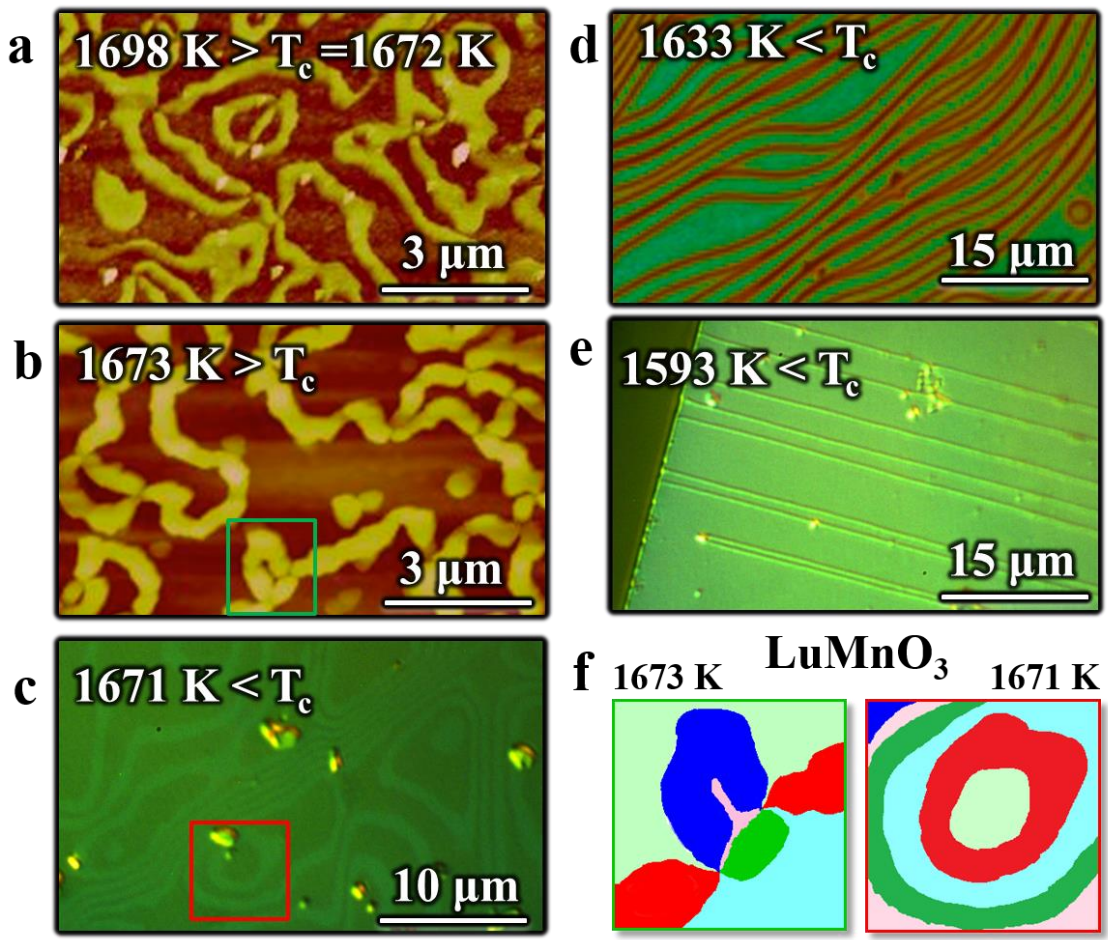

**Fig. 4**

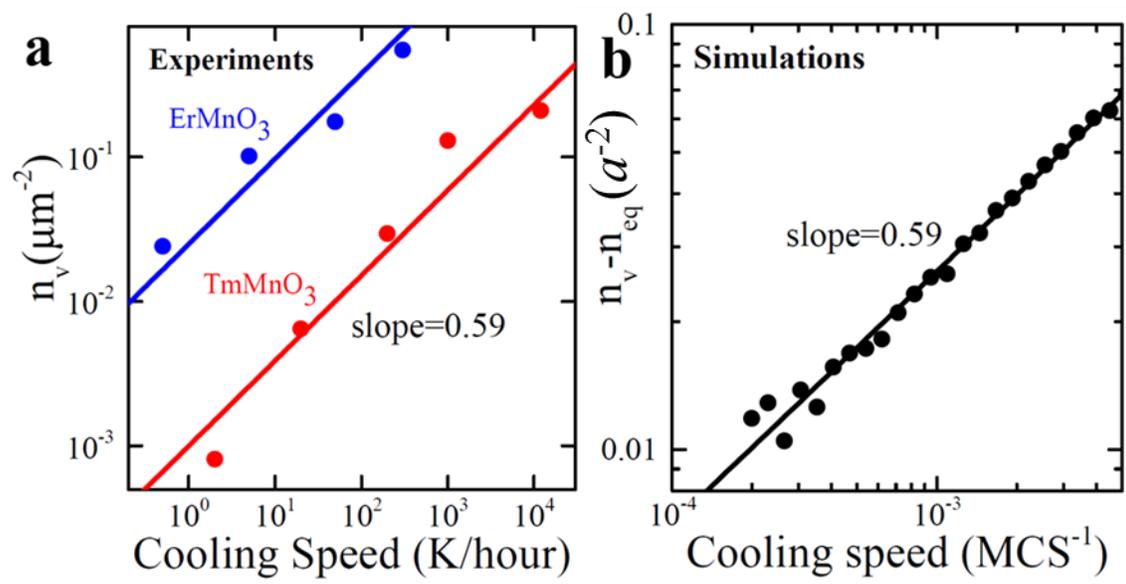

Fig. 5

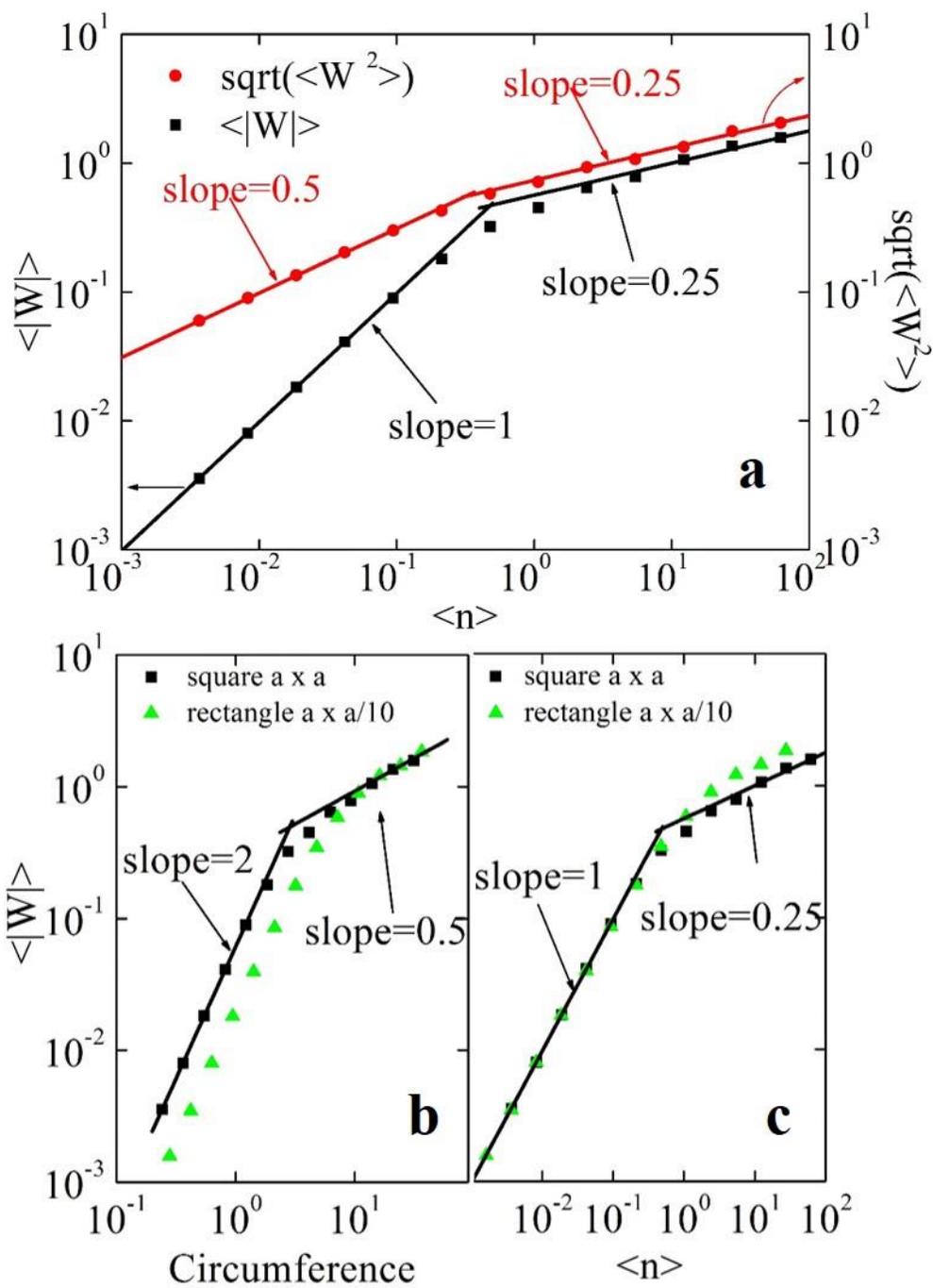

**Fig. 6**

# Duality

The local order parameter of our problem is a complex field $\phi_j = |\phi_j|e^{i\varphi_j}$, that takes six possible values ($Z_2 \times Z_3$) corresponding to the even times of a clock. By assuming that the local trimerization and dipole moments develop above $T_c$, we neglect the amplitude ($|\phi_j|$) fluctuations near $T_c$. The six orientations of $\phi_j$ are enforced by an effective potential, $V(\phi_j)=A\cos(6\varphi_j)$, which reflects the anisotropy of the underlying crystal lattice. $V(\phi_j)$ is dangerously irrelevant at $T_c$, i.e., the coarse grained action near the ferroelectric transition becomes identical to the isotropic $\phi^4$ action for the normal to superfluid transition of a neutral system like $^4$He [see Fig.1(a)]:

$$H_\phi = m_\phi^2 \phi^2 + u_\phi \phi^4 + (\nabla \phi)^2 \qquad (2)$$

The superfluid to normal transition occurs via proliferation of vortex lines at $T>T_c$. The problem admits a dual description, in which the proliferation of vortex lines spanning the whole system arises from a condensation of a *dual or "disorder" field* $\psi=|\psi|e^{i\theta}$ minimally coupled to an effective gauge field. Below $T_c$, the "photon" of this gauge field is the Goldstone mode of the superfluid field $\phi$. This photon acquires a finite mass via the Higgs mechanism for $T>T_c$ [see Fig. 1(b)]. Consequently, the Biot-Savart (Coulomb) interaction between vortex segments for $T<T_c$ becomes screened (Yukawa) for $T>T_c$.

The dual description is obtained after a sequence of transformations. The original $\phi^4$ theory (B1) for a *neutral* superfluid is first mapped into a loop gas of vortices coupled to a vector gauge field **A** generated by the smooth phase fluctuations of the original field $\phi$. The fluctuating vortex loops are then described by a disorder $|\psi|^4$ field theory in which the vortex loops correspond to "supercurrents" of $\psi$, which remain minimally coupled to $\mathbf{A}$[14-16]:

$$H_\psi = m_\psi^2 \psi^2 + u_\psi \psi^4 + \frac{1}{2t}\left|(\nabla - iq_{eff}\mathbf{A})\psi\right|^2 + \frac{1}{2}(\nabla \times \mathbf{A})^2$$

The constants $t$ and $q_{eff}$ are determined by non-universal parameters, such as the vortex core energy and the transition temperature[14-16]. Having direct experimental access to the vortex field, we can observe the Higgs condensation of $\psi$: the emergence of vortex lines that span the whole system above $T_c$ implies that superfluid currents of the *disorder field $\psi$* connect opposite ends of the sample, i.e., the disorder field has condensed into a "superfluid state" [see Fig.1(b)].

**Box 1**

# Supplementary Information

# Topological defects as relics of emergent continuous symmetry and Higgs condensation of disorder in ferroelectrics

1. Methods

Plate-like RMnO$_3$ single crystals with flat surfaces perpendicular to the c-axis were grown using a flux method with Bi$_2$O$_3$ flux (see Fig. S1). One LuMnO$_3$ crystal was annealed at 1723 K to induce vortex domains, and then was etched in phosphoric acid at 150 °C for 30 minutes to reveal the domain pattern on the top surface. Afterwards, the top surface was polished, and then a Piezo-response Force Microscope (PFM) image was taken on the polished surface. We repeated the polishing and taking PFM image a few times. PFM and Atomic Force Microscope (AFM) images were obtained using a Nanoscope IIIA (Veeco).

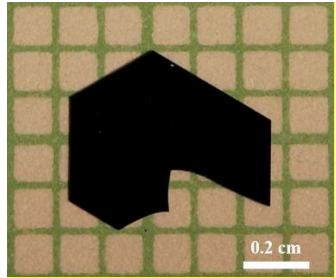

Figure S1: A Plate-like single crystal with a flat a-b surface and hexagonal edges

2. Simulation details

The positions of the vortex and anti-vortex cores are obtained by calculating the total phase difference $\Delta\phi$ along a unit cell. The core is located at the unit cell with $\Delta\phi=\pm 2\pi$. We neglect the possibility of vortices with absolute value of the winding number larger than one because their energy grows as the square of the winding number, i.e., they are exponentially suppressed near $T_c$.

Fig. 5 b shows the vortex density obtained after a simulated annealing of Eq. (1) on a lattice of L×L×$L_z$=96$a$×96$a$×96$a$ sites from the equilibrium state at $T_i$ =6.0 $J$ to the final state at $T_f$≈ 0.92 $T_c$. Here $a$ is the lattice parameter. 200 independent runs were performed to obtain a good statistics. Besides the vortices generated according to Kibble-Zurek (KZ) mechanism, the thermally excited vortices $n_{eq}$ are abundant near $T_c$. Then, the density of thermal vortices, $n_{eq}$, must be subtracted to obtain the scaling predicted by the KZ mechanism. The choice of $T_f$ ≈ 0.92 $T_c$ is a good compromise to avoid two undesirable effects. On the one hand, the relaxational dynamics that leads to vortices-antivortex annihilation becomes important at low enough temperatures. On the other hand, close enough to $T_c$ it becomes very difficult to obtain a reliable estimate of $n_{eq}$ at $T_f$ because of the well-known critical slowing down. The exponent $\zeta=2\nu/(1+z\nu)$ is obtained by fitting the final vortex density, $n_v$, as a function of the cooling rate $r$ with the function $n_v(r)=n_{eq}+r^\zeta$. The value $n_{eq}$≈0.121 that results from the fitting is consistent with the value $n_{eq}$≈0.105(22) obtained from simulations in equilibrium.

To understand the asymmetry between the annealing from above $T_c$ and below $T_c$, we took snapshots during the annealing process. Several typical configurations for both $T_i$ =3.2$J$ >$T_c$ and $T_i$ =2.9 $J$<$T_c$ are shown in Fig. S2. As expected, only small vortex loops exist for $T_i$ =2.9 $J$<$T_c$. These loops shrink and disappear quickly during the annealing process (c.f. Fig. S2 (e)) leading to a single domain final state (c.f. Fig. S2 (f)) (we are not including long-range dipolar interactions in our model). In contrast, vortex lines that span the whole system appear for $T_i$ =3.2$J$ >$T_c$, in addition to the small vortex loops. Once again, the small loops quickly disappear along the annealing process (c.f. Fig. S2 (b)). Nevertheless, the percolating vortex lines remain because they are well separated and the interaction between them is screened by the presence of small vortex loops (c.f. Figs. S2 (c) and (g)).

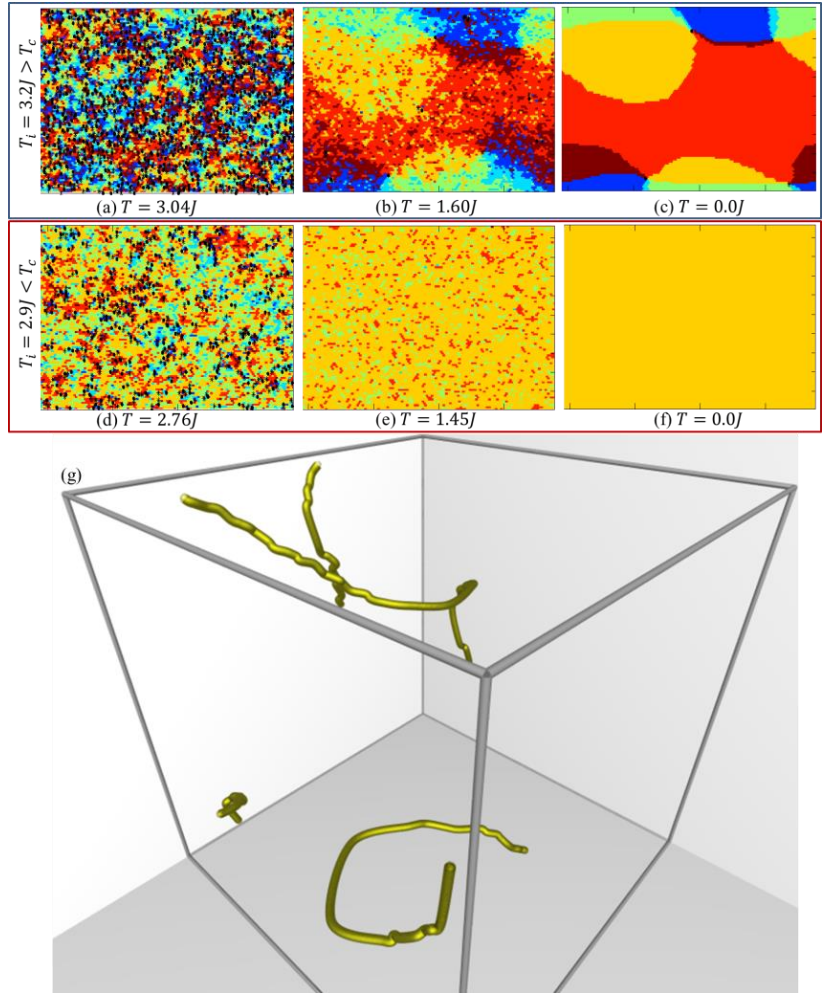

**FIG. S2** (a)-(f) Typical domain and vortex configurations obtained during the annealing process from a state above (a-c) and below $T_c$ (d-f). The different colors denote the six different domains. Dots (crosses) represent vortices (antivortices). (g) The corresponding 3D plot of (c). Note these vortex lines are connected because of the periodic boundary conditions. The lattice size is L×L×$L_z$=96$a$×96$a$×96$a$ and the annealing rate is r=0.001per Monte Carlo Sweep.

3. **Optical images of samples used for vortex-antivortex correlation function analysis**

Large-region optical images have been taken on ErMnO$_3$, YMnO$_3$ and YbMnO$_3$ (green, black, and red data in Fig. 3(b), respectively) after chemical etching. Coordinates for every vortex and anti-vortex core were found and then analyzed for vortex-antivortex pair correlation function. The average distance between the defects for all the 3 samples: 2.3 μm for YMnO$_3$, 2.5 μm for YbMnO$_3$, and 14.8 μm for ErMnO$_3$. The winding numbers were calculated using the coordinates of the vortices and antivortices in these optical images. Contours of square and rectangular shapes with various sizes were used in the calculations. The results for all three samples are very similar. Fig. 6 shows the winding number data for the YMnO$_3$ sample, for which the largest number of the defects (~4100) was recorded.

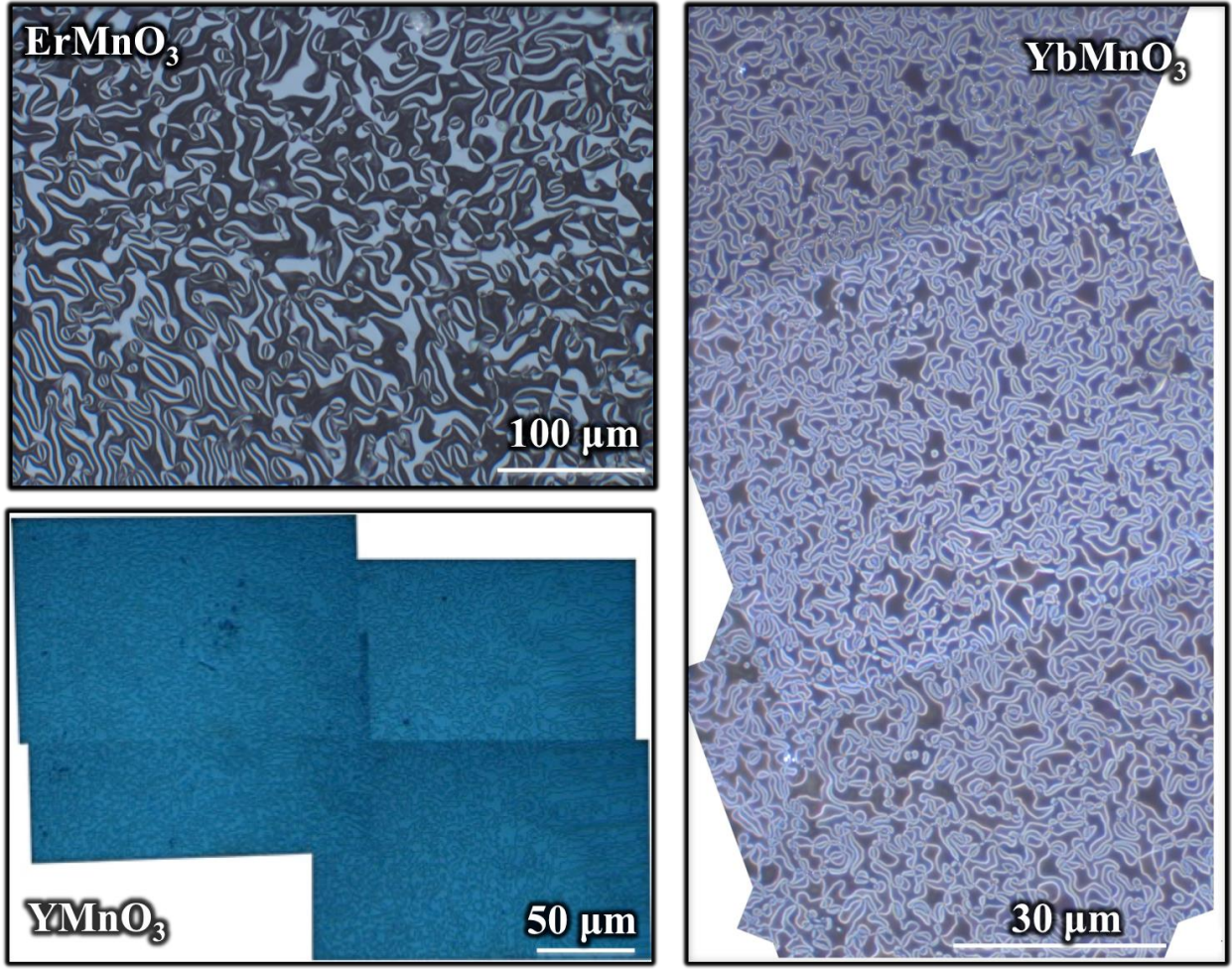

**Figure S3:** Optical images of ErMnO$_3$, YMnO$_3$ and YbMnO$_3$ (green, black, and red data in Fig. 3(b), respectively), on which vortex-antivortex pair correlation function analysis has been performed. The average distance between the defects for all the 3 samples: 2.3 μm for YMnO$_3$, 2.5 μm for YbMnO$_3$, and 14.8 μm for ErMnO$_3$.

4. **Optical images of TmMnO$_3$ of different annealing temperature**

A series of annealing with different cooling rates have been done: 2, 20, 200, 1000 and 12000 K/h. Optical images of 4 of them are shown in Figure below. Numbers of vortices and antivortices were counted.

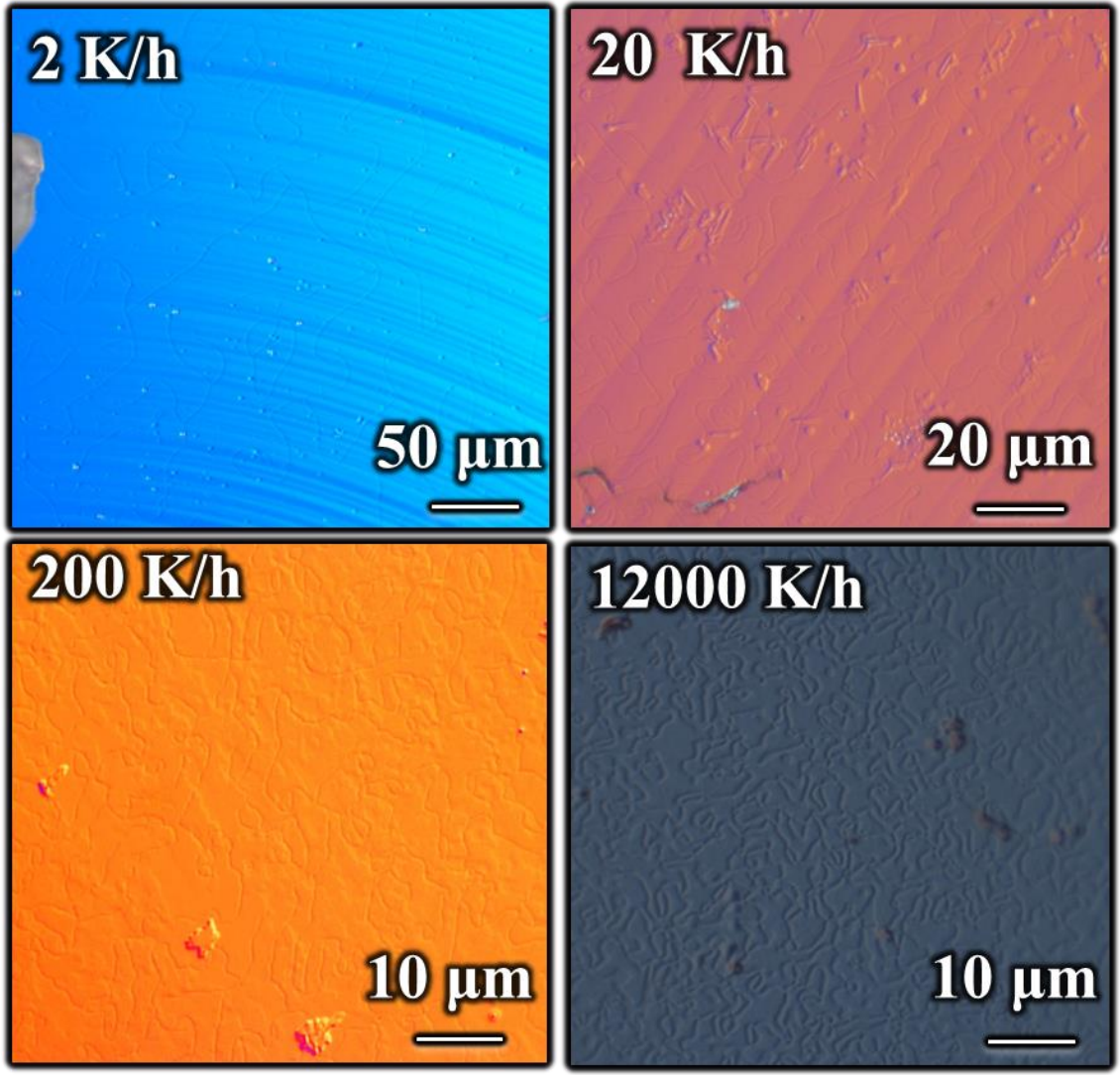

**Figure S4:** Optical images of TmMnO$_3$ with 4 different cooling rates after chemical etching are shown: Cooling rate are 2, 20, 200, 1000 and 12000 K/h (calibrated from furnace cooling) cross the ferroelectric transition temperature, which is 1523 K for TmMnO$_3$.